\documentclass{elsart}
\usepackage{graphicx}
\usepackage{amsmath}
\newcommand{\lleq}[1]{\label{#1} }
\def\bp {{ \mathbf{p} }} 
\def\bq {{ \mathbf{q} }} 
\def\bz {{ \mathbf{z} }}

\begin{document}
\begin{frontmatter}

\title{\Large Transverse-longitudinal HBT correlations\\ 
  in $\mathbf{p\bar p}$ collisions at $\mathbf{\sqrt{s} = 630}$ GeV}

\author[STELLENBOSCH]{H.C.\ Eggers,}
\author[HEPHY]       {B.\ Buschbeck} and
\author[IMT,STELLENBOSCH]{F.J.\ October}

\address[STELLENBOSCH]{Department of Physics, 
  University of Stellenbosch,\\
  7602 Stellenbosch, South Africa}
\address[HEPHY]{Institut f\"ur Hochenergiephysik,
  Nikolsdorfergasse 18, A--1050 Vienna, Austria}
\address[IMT]{Institute for Maritime Technology, 
  P.O.\ Box 181, 7995 Simonstown, South Africa}

\begin{abstract}
  Correlations of like-sign pion pairs emerging from proton-antiproton
  collisions are analysed in the two-dimensional $(q_L,q_T)$
  decomposition of the three-momentum difference $\bq$. While the
  data cannot be adequately represented by gaussian, exponential,
  power-law or Edgeworth parametrisations, more elaborate ones such as
  L\'evy and an exponential with a cross term do better. A two-scale
  model using a hard cut to separate small and large scales may
  indicate a core that is more prolate than the halo. Consideration
  not only of the interference peak at small $(q_L,q_T)$, but also of
  the shape of the correlation distribution at intermediate momentum
  differences is crucial to understanding the data.

  \noindent
  PACS: 13.85.Hd, 13.87.Fh, 13.85.-t, 25.75.Gz
\end{abstract}

\end{frontmatter}

\section{Introduction}

The measurement of correlations between identical final-state
particles, also commonly termed ``Bose-Einstein correlations'' or
``HBT effect'' after the fathers of intensity interferometry in
astronomy, has become a valuable tool in the quest to understand the
spacetime structure of high-energy collision processes
\cite{Hei99a,Cso00a}. The recently-coined term of \textit{femtoscopy}
highlights the possibilities of unraveling the interplay of kinematics
and dynamics at the femtometer level based on the symmetrisation of
identical bosons.

The current concentrated effort at RHIC to quantify and understand
ultrarelativistic nuclear collisions \cite{Lis05b} relies extensively
on comparisons with baseline scenarios constructed from the
corresponding ``trivial'' hadron-hadron sample. In this context, the
UA1 experiment continues to be relevant and interesting, even though
data-taking at the CERN SPS has long ceased.  Current experimental
energies of $200$ AGeV at RHIC are still below those available to UA1
by a factor three, so that our results may also provide a window on
possible energy dependencies of current investigations.

One of the relevant theoretical frameworks is constructed in terms of
a three-dimensional decomposition of the pair momentum difference,
$\bq = \bp_1 - \bp_2$ along the collision axis, the pair transverse
momentum and the ``side'' direction orthogonal to both \cite{Ber88a}.
In this letter, we provide results on HBT analysis in terms of the
simpler two-dimensional decomposition, defining in the usual way $q_L
= |\mathbf{q}_L| = |(\mathbf{q}\mathbf{\cdot}\hat\bz)\, \hat\bz|$,
with $\hat\bz$ the beam direction, and $q_T = |\mathbf{q} -
\mathbf{q}_L|$. We analyse correlations between like-sign (LS) pion
pairs in terms of the normalised second moment,
\begin{equation} 
\label{iac}
R_2(q_L,q_T) = \frac{\rho_2^{\rm sib}(q_L,q_T)}
                    {\rho_2^{\rm ref}(q_L,q_T)}\,,
\end{equation}
where $\rho_2^{\rm sib}(q_L,q_T)$ counts ``sibling'' LS pairs 
from the same event, while $\rho_2^{\rm ref}(q_L,q_T)$
counts ``reference'' pairs made up through event mixing.

\section{Data sample, cuts and corrections}

Like-sign pion pairs from approximately 2.45 million minimum-bias
events \cite{UA1-85} measured by the UA1 central detector in 1985 and
1987 were analysed. We applied the same single-track cuts used
previously \cite{Bus00a}, including $p_\bot \geq 0.15$~GeV $| \eta |
\leq 3$ and $45^\circ \leq |\phi| \leq 135^\circ$. Good measurement
quality and fitted track length $\ell > 30$ cm were required.  The
sample contains mainly pions with an estimated 15\% contamination 
of charged kaons and protons
\cite{UA1-92a}.

As in previous work \cite{Egg97a,Bus00a}, an angle cut was
applied due to limited angular resolution in the central detector
\cite{Lip90a}, and pairs were required to have minimum four-momentum
difference $Q^2 = \sqrt{-(p_1-p_2)^2} > 0.0003$ GeV$^2$.

Spurious ``split--track'' pairs strongly influence pair counts at
small relative momenta. Besides eliminating these pairs with the same
algorithm used in previous analyses, we now also correct for the fact
that this necessarily eliminates some physical like-sign track pairs.
We determine correction factors for each $(q_L,q_T)$ bin measured in
the detector rest system and charged-multiplicity subsample (see
below) by passing unlike-sign pairs through the same split-track
algorithm. These correction factors are substantial at small
$(q_L,q_T)$, reaching 1.9 for low-multiplicity subsamples. Errors
shown and quoted take into account the additional uncertainty
introduced by the correction.

We corrected for Coulomb repulsion by parametrising the Bowler Coulomb
correction in the invariant momentum difference \cite{Bow91a} with an
exponentially damped Gamov factor $G(Q)$ \cite{Bri95a}, $ F_{\rm
  coul}(Q) = 1 + \left[ G(Q) - 1 \right] \, \exp(-Q/Q_{\rm eff})$,
with a best-fit value $Q_{\rm eff} = 0.173 \pm 0.001$ GeV.

Correlations from subsamples of fixed $N$, the observed charged
multiplicity in the whole azimuth, were summed separately in the
numerator and denominator
\begin{equation} 
\label{dsc}
R_2(\mathbf{q}) = 
\frac
{\sum_N P_N\, \rho_2(\bq\,|\,N)}
{\sum_N P_N\, (1-N^{-1})\,
\rho_1{\otimes}\rho_1(\bq\,|\,N)}
\,.
\end{equation}
The sum $\sum_N P_N$ was implemented in terms of ten subsamples
defined by multiplicity $N$. The reference for each $N$-subsample,
$\rho_1{\otimes}\rho_1(\bq\,|\,N)$, was constructed by creating for
each sibling event a set of fake events with the same multiplicity,
using tracks randomly selected from the corresponding subsample track
pool. The factor $(1-N^{-1}) = N(N{-}1)/N^2$ arises because the
uncorrelated reference for fixed-$N$ subsamples is not the poisson but
the multinomial distribution \cite{Lip96a}. All HBT quantities are
measured in the longitudinal co-moving system (LCMS) of the pion pair.

\section{Parametrisations}

We use the generic parametrisation $ R_2(q_L,q_T) = \gamma \left[ 1 +
  \lambda \left| S \right|^2 \right], $ where the multiplicative
parameter $\gamma$ corrects for any remaining effect of the overall
multiplicity distribution, $\lambda$ is the ``chaoticity parameter''
and $S(q_L,q_T)$ corresponds to the Fourier-transformed source
function. For $|S|^2$, we implemented the simple gaussian, the
gaussian with cross-term \cite{Cha95a}, the simple exponential, and
the exponential with cross-term parametrisations,
\begin{eqnarray}
\lleq{prc}
|S|^2 &=& \exp(-R_L^2 q_L^2 - R_T^2 q_T^2) \,,\\
\lleq{prd}
|S|^2 &=& \exp(-R_L^2 q_L^2 - R_T^2 q_T^2 - 2 R_{LT}^2 q_L q_T)\,, \\
\lleq{pre}
|S|^2 &=& \exp(-R_L q_L - R_T q_T) \,,\\
\lleq{prf}
|S|^2 &=& \exp(-R_L q_L - R_T q_T - 2 R_{LT} \sqrt{q_L\, q_T})
\,,
\end{eqnarray}
as well as a power-law parametrisation,
\begin{equation}
\lleq{pri}
R_2 = \gamma \left[ 1 + (R_L q_L)^{-\alpha_L} \, (R_T q_T)^{-\alpha_T} 
            \right]\,.
\end{equation}
Apart from the above parametrisations, nongaussian distributions can
also be approached through Edgeworth expansions\footnote{Third-order
  cumulants $\kappa_{3,d}$ must be zero due to the symmetry of
  $R_2(q_L,q_T)$.  } \cite{Cso93a,STAR05a},
\begin{equation} 
\lleq{prj}
\left| S \right|^2 = \exp(-R_L^2 q_L^2 - R_T^2 q_T^2)
\prod_{d=L,T}\left[ 
1 + \kappa_{4,d}\; H_4(\sqrt{2}\,R_d\,q_d)/24 \right],
\end{equation}
or through L\'evy distributions \cite{Cso04a},
\begin{eqnarray} 
  \lleq{prk}
  \left| S \right|^2 &=& \exp(-R_L^2 q_L^2 - R_T^2 q_T^2)^{\alpha/2}
  \,,\\
  \lleq{prl}
  \left| S \right|^2 &=& \exp(-R_L^2 q_L^2 - R_T^2 q_T^2 -
  2 R_{LT}^2 q_L q_T)^{\alpha/2}.
\end{eqnarray}
All fits with the above parametrisations were performed over the whole
region $0 \leq q_L,q_T \leq 0.5\,$GeV but omitting the bin $(q_L,q_T)
< (0.02,0.02)\,$GeV which suffers from multiple large corrections and
detector effects. Note that values for $R_L$ and $R_T$ cannot be
compared between different parame\-trisations; note also that these
parameters play the traditional role of ``radius'' of a distribution,
and can be related to various source parameters \cite{Cha95b}, only
for the gaussian cases (\ref{prc})--(\ref{prd}).

\section{Analysis and Results}

\subsection{The region of small $(q_L,q_T)$}

Fig.~1 shows slices of $R_2(q_L,q_T)$ for fixed $q_T$ bins in the
upper and fixed $q_L$ bins in the lower panels, together with the
global fits based on Eqs.~(\ref{prc})--(\ref{prl}). The Gauss,
Gauss-with-cross-term, and Edgeworth fits are practically
indistinguishable, as $R_{LT}$ in (\ref{prd}) is compatible with zero
and cumulants $\kappa_{4,L}$ and $\kappa_{4,T}$ in (\ref{prj}) do not
improve the fit.  All three are equally bad, with $\chi^2/{\rm NDF}
\sim 3.3$.  The simple exponential (\ref{pre}) (not shown) is even
worse at 4.0. The cross-term exponential, however, has
$\chi^2/\mbox{NDF} = 1.7$. The power law (\ref{pri}) is a disaster at
$\chi^2/\mbox{NDF} = 16$.

Both L\'evy-based parametrisations (\ref{prk})--(\ref{prl}) fare
better in reproducing the strong peak observed in the data, with
$\chi^2/\mbox{NDF} = 1.45$ and $1.13$ respectively. However, the fit
parameters $\lambda$, $R_L$, $R_T$, $R_{LT}$ and $\alpha$ are strongly
correlated so that there is no unique minimum and therefore no unique
set of best-fit parameter values. All that can be said with some
confidence is that $\alpha$ is around 0.20--0.23, that $R_L$ and $R_T$
are of order $10^3\,$fm with $R_L/R_T$ = 1.6, and that $R_{LT}^2$ is
negative; we will return to this below. Omitting a second
small-$(q_L,q_T)$ bin from the L\'evy fits renders them even more
unstable. This is hardly surprising, as the above parameters
collectively depend strongly on the exact shape of the peak in the
very small $(q_L,q_T)$ region, the very region that experimental
measurement struggles to resolve.

\subsection{Observations at intermediate $(q_L,q_T)$}

Fig.~1, revealing as it is, tells only part of the story as it focuses
exclusively on the peak at small $|\bq|$. Intermediate scales, it
turns out, are very important.

In Fig.~2, the same data and some of the fits are shown as colour maps
over the full set of bins, but with the peaks in the lower left
corners truncated at $R_2 = 1.9$ in order to highlight structure at
intermediate $(q_L,q_T)$ scales. Panel (b) shows that the simple
exponential (\ref{pre}) cannot possibly describe the data as its shape
at intermediate scales is linear, while the data in (a) is elliptic.
The L\'evy without cross term (\ref{prk}), Panel (e), matches the
shape better. Panels (c) and (f), representing the exponential and
L\'evy with cross terms included, most accurately reflect the data
structure. The gaussian forms (\ref{prd}) and (\ref{prj}) have a shape
similar to that in (e). From Panel (d), it is immediately apparent
that the spectacular failure of the power law (\ref{pri}) stems from
its hyperbolic shape.  A ``sum power law'' $R_2 = \gamma\left[1 +
  (R_L\, q_L)^{-\alpha_L} + (R_T\, q_T)^{-\alpha_T} \right]$ fares
even worse.

Figure 2 also makes clear that the source --- whatever its exact shape
--- is prolate, i.e.\ $R_L > R_T$ must hold for all parametrisations.

\subsection{A two-scale approach.}

The strong peak seen in the data and the comparatively large values
for $\chi^2$ for the above parametrisations suggests that there may be
two scales in the system. Borrowing the terms ``core'' and ``halo'' 
from the literature \cite{Cso94a}, 
one may try a simultaneous fit to a double-gaussian
parametrisation \cite{Led79a,NA22-93a},
\begin{equation} \label{rsg}
R_2(q_L,q_T) = \gamma 
\left[ 1 
   + \lambda_C \exp(-R_{LC}^2 q_L^2 - R_{TC}^2 q_T^2 )
   + \lambda_H \exp(-R_{LH}^2 q_L^2 - R_{TH}^2 q_T^2 )
\right].
\end{equation}
However, again there are too many parameters, so that there is no
unique set of best-fit values. For this reason, we try a ``two-scale''
procedure, starting from the assumption that the two scales implicit
in (\ref{rsg}) can be separated by a hard cut in the data.  First, we
fit to the core gaussian only bins with momentum differences larger
than a cutoff ($q_L> q_{\rm cut}$ or $q_T > q_{\rm cut}$), thereby
fixing $\gamma$, $\lambda_C, R_{LC}$ and $R_{TC}$.  This core gaussian
is then subtracted from all data, after which the remaining halo
``data'' with $(q_L,q_T) \leq (q_{\rm cut},q_{\rm cut})$ is fit with
the halo gaussian $\gamma\,\lambda_H \exp(-R_{LH}^2 q_L^2 - R_{TH}^2
q_T^2)$ for fixed $\gamma$. Finally, the resulting core and halo fits
are combined.

This procedure must be tested for consistency. As shown in Fig.~3(a),
the joint $\chi^2$ for both fits as a function of $q_{\rm cut} =
0.02\,n_{\rm bin}$ is found to have a minimum in the interval $q_{\rm
  cut} =$ 180--240~MeV. The joint best $\chi^2/{\rm NDF} = 1.28$ for
the two-scale model is comparable to those for the L\'evy fits and fit
values are stable, so that numbers can be quoted. The $R_L$, $R_T$,
$\lambda$ and $R_L/R_T$ values corresponding to the four smallest
$\chi^2$ values are shown as filled points in Fig.~3(b)--(e).
Averaging these numbers, we estimate
$R_{LC} = 0.75\pm 0.02$~fm, %
$R_{TC} = 0.45\pm 0.02$~fm, %
$R_{LH} = 2.26\pm 0.07$~fm, and %
$R_{TH} = 1.89\pm 0.10$~fm, %
signalling a prolate core and a somewhat more spherical halo.  We note
that best-fit chaoticities $\lambda_C = 0.23 \pm 0.01$ and $\lambda_H
= 0.92 \pm 0.04$ are below the theoretical limit of 1, while the large
intercept seen in the data itself ($R_2(0,0) > 3.7$) and corresponding
single-component chaoticities ($\lambda \sim 2.7$) violate this limit.

In Fig.~4, the upper lines show the combined two-scale best fit for
$q_T$ and $q_L$ slices, with the lower lines corresponding to the
core. While there is substantial improvement over simple
parametrisations especially at intermediate scales, data points in the
peak remain consistently above even this fit.

It must be emphasised that this two-scale procedure can work only if
its assumptions are confirmed \textit{a posteriori.}  First, the
resulting halo gaussian must, and does, become negligible at scales
larger than $q_{\rm cut}$ for the ansatz to be valid. Also, the clear
separation between the sizes of the core and halo radii seen in Fig.~3
do not contradict the assumption of the presence of two scales. For
this data set, the two-scale model is consistent.

\subsection{Systematic errors and uncorrected data}

The above results have shown that the data appears to have a strong
peak below 0.10 GeV which significantly exceeds all parametrisations
tried. While it is tempting to conclude that the peak represents some
physical effect, other possibilities must be checked. We hence
conducted a survey of effects that various cuts and corrections have
on $R_2$. We find that the angle cut and $Q^2$ cut and the restriction
in azimuth have very little effect on the normalised moment and that
any systematic error due to these is of the order of a few percent.

However, the correction for the unwarranted removal of real LS track
pairs by the split-track algorithm is large, and it is concentrated in
precisely the region where the large peak occurs. Naturally, one must
ask whether the entire excess of the data over various
parametrisations reflects nothing but the correction itself. Taking
the extreme approach of leaving the correction out altogether, we have
repeated the entire analysis for the uncorrected data, which is shown
in Fig.~5 together with the same set of parametrisations used in
Fig.~1 plus the power law (\ref{pri}).

It is immediately apparent that the uncorrected data peak cannot be
described by gaussian parametrisations either ($\chi^2/{\rm NDF} \sim
3.5$). With regard to shape (not shown), the general elliptic and
prolate form of the data remains unchanged. The various
parametrisations follow the pattern set by Fig.~2: again, the simple
exponential and power law fail because they do not reproduce the shape
of the data; again, the L\'evy and exponential with cross term work
better, with the latter faring best. While fit parameter values (e.g.
of the L\'evy-fits with $R_L \sim$ 2.5--2.9 fm, $R_T \sim$ 1.6--1.8 fm
and $\alpha \sim 0.7$) differ from those of the corrected data, the
conclusions regarding the success of particular parametrisations are
hence independent of the split-track correction.

\section{Discussion and Conclusions}

The present results on HBT correlations extend UA1 observations to
multidimensional correlations for the first time, which, at 630 GeV,
represent the highest CMS energy at which this has been done. Two
issues stand out: the strength of the peak at small momentum
differences, and the importance of shape at intermediate scales.

With a strong peak at small $q_L,q_T \le 0.1\,$GeV, our high-statistic
data, with and without corrections, rules out parametrisations based
on single gaussians and their derivatives.  While some other
parametrisations are strongly peaked, none of those tried, not even
the L\'evy and two-scale cases, reproduces the peak data convincingly.
This may hint that ``something else is going on'', be it the influence
of jets, resonances, clustering effects or some other unknown factor.
The two-scale results may hint that the ``halo'' could be due to
short-lived resonances such as the $\rho$.

We stress that it is unlikely that the strong peak seen in our data at
small momentum differences is due to bias. First, the peak persists
even without the Coulomb or split-track corrections.  Second, the
nongaussian behaviour seen in this paper is in line with UA1
one-dimensional correlation structures seen earlier in the form of
nonzero higher-order cumulants \cite{UA1-92a,Car90d} and in the power
law in $Q^2$ \cite{Egg97a,UA1-93a}, even though the data had not been
corrected in the way it has been here. Third, the presence of
unidentified kaons and protons in the sample imply that the real peak
should exceed the one shown here, so that the present results are
conservative. It should be noted that a number of other hadronic 
experiments \cite{AFS-83a,NA22-95a,NA22-96a} have also seen
significant deviations from gaussian behaviour at small $|\mathbf{q}|$.

Besides the structure of the peak seen at small scales, the shape of
the distribution at intermediate scales is seen to provide valuable
additional information. Fig.~2 shows that the data has without doubt
an elliptic and prolate shape, the latter in the sense that the
distribution in momentum space is narrower in the longitudinal than in
the transverse direction. This is confirmed also within the two-scale
method and for the uncorrected data. Significantly, these conclusions
are independent of the various parametrisations and corrections.

Second, plots of shape (whether in colour or as contour lines) help to
constrain possible parametrisations: whereas in the one-dimensional
analysis of UA1 data \cite{Egg97a}, the exponential and power-law
parametrisations were found to be superior to the gaussian, their
extension to two dimensions fails badly because their contour lines
are straight lines or even hyperbolic.

Considering shape in terms of different parametrisations is
complementary to decompositions into cartesian and spherical harmonics
\cite{Dan05a,Lis05a}, which can be expected to work best for
near-gaussian data. We note, however, that the interrelationship
between shape and algebraic form of a parametrisation is less than
obvious: elliptic shape is exhibited not only by bilinear forms such
as Eqs.~(\ref{prc})--(\ref{prd}) and (\ref{prk})--(\ref{prl}) but
also, surprisingly, by the exponential-with-cross-term, whose success
depends strongly on the sign of the cross term. For both exponential
(\ref{prf}) and L\'evy with cross-term (\ref{prl}), our data clearly
prefers \textit{negative} values for the cross term ($R_{LT}$ and
$R_{LT}^2$ respectively); positive values result in more hyperbolic
shapes, in conflict with the data.

Their fit instability nonwithstanding, the L\'evy parametrisations,
which have the right shape and an acceptable peak, appear promising.
Taking these at face value would imply that the corresponding source
falls off with a power-law tail at large distances
\cite{Cso04a,Bial,Uty}.


\section*{Acknowledgments}

\noindent
We thank A.\ Bia{\l}as, T.\ Cs\"org\H{o}, K.\ Fia{\l}kowski and W.\
Kittel for useful discussions, and the UA1 collaboration for providing
the data. Technical support by G.\ Walzel is gratefully acknowledged.
HCE and BB thank respectively the Institute for High Energy Physics in
Vienna, CENPA at the University of Washington and the Department of
Physics in Stellenbosch for kind hospitality.  This work was funded in
part by the South African National Research Foundation.


\vspace*{20mm}

\begin{figure}[htb]
\includegraphics[width=130mm,bb=55 30 530 200,clip=] 
    {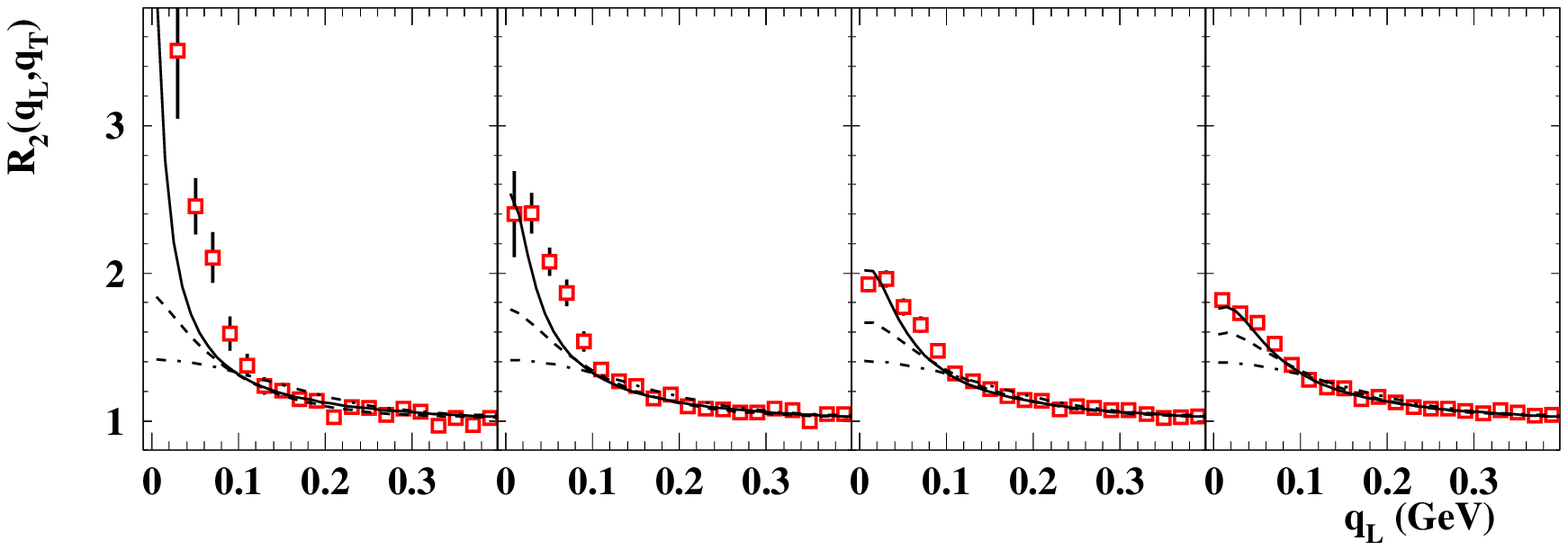}\\[1mm]
\includegraphics[width=130mm,bb=55 30 530 200,clip=] 
    {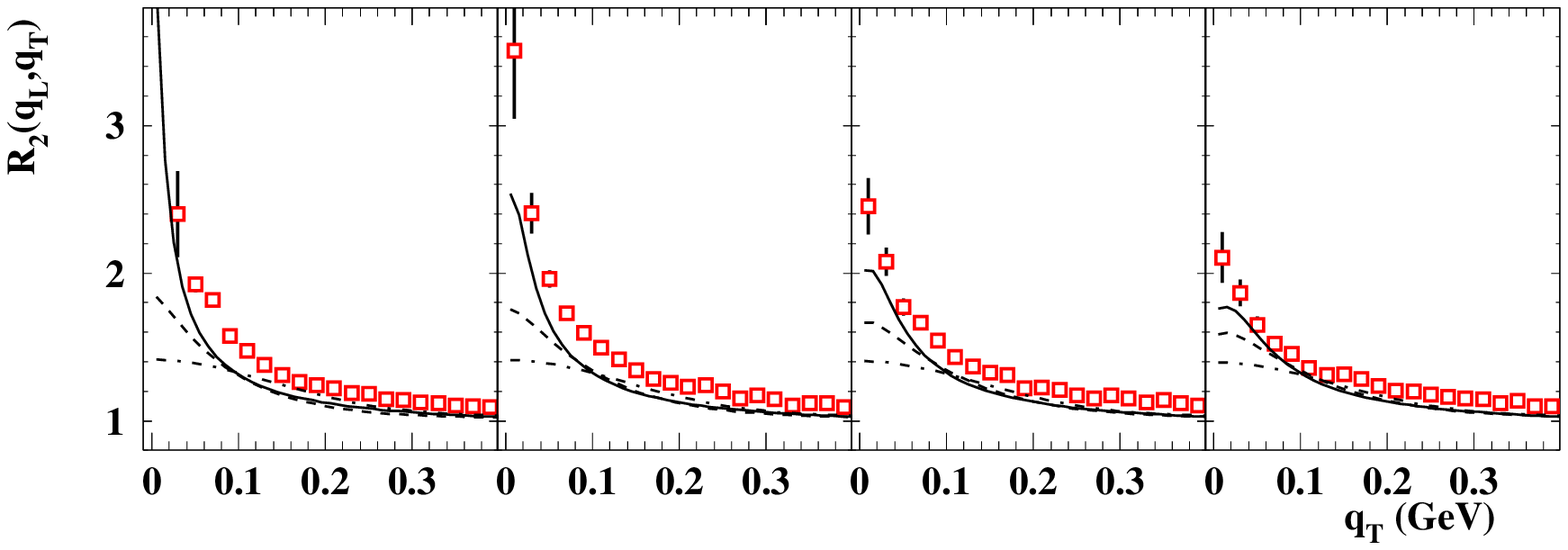}\\[1mm]
    \caption{Upper panels: $R_2(q_L,q_T)$ data and best fits, shown
      left to right for slices with fixed $q_T = $ 0.00--0.02,
      0.02--0.04, 0.04--0.06 and 0.06--0.08 GeV bins. Solid lines:
      L\'evy fit (\ref{prl}); dashed: exponential with cross term;
      dash-dotted: Gauss/Edgeworth.  Lower panels: $R_2(q_L,q_T)$ and
      the same fits for corresponding fixed-$q_L$ slices. Throughout
      this letter, fits are performed over all 624 data points, not
      just the data points, slices and intervals shown in the
      figures.}
\end{figure}
\begin{figure}[htb]
\label{fshape}
\begin{center}
\includegraphics[width=130mm,bb=50 30 600 420,clip=] 
    {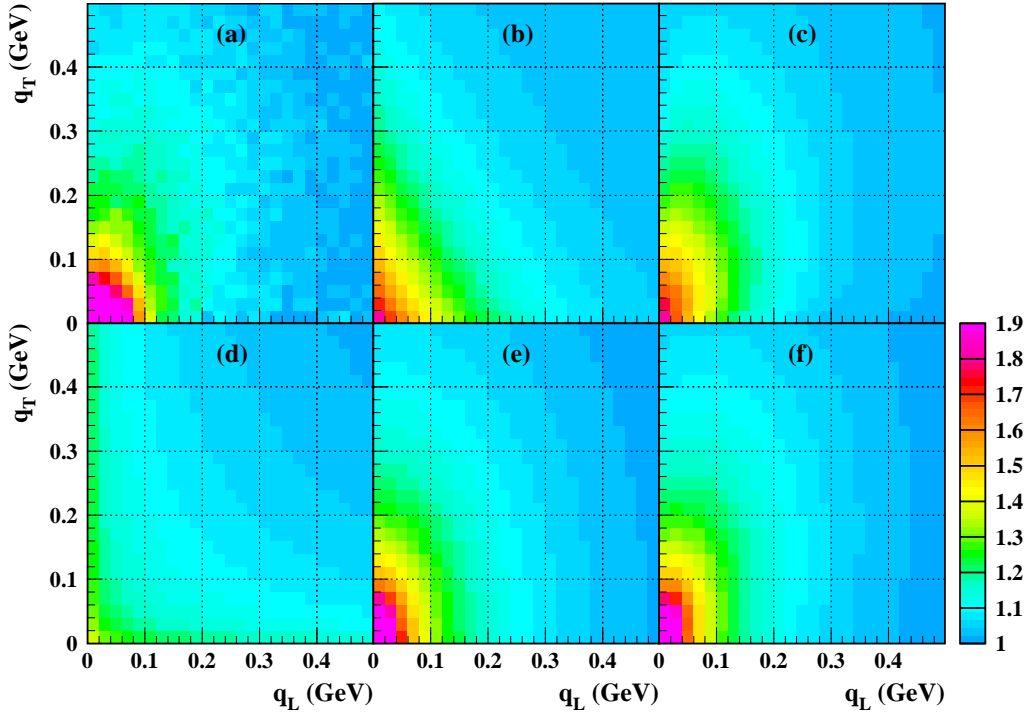}\\[1mm]
\end{center}
\caption{Shapes of $R_2(q_L,q_T)$ data and fits at intermediate
  scales. Panels show (a) UA1 data, (b) exponential, (c) exponential
  with cross term (d) power-law, (e) L\'evy, (f) L\'evy with cross
  term.  Plots are truncated vertically to $R_2 \leq 1.9$ to highlight
  structure at intermediate scales.}
  \ \\[10mm]
\end{figure}
\begin{figure}[htb]
\begin{center}
\includegraphics[width=135mm,bb=40 30 540 210,clip=] 
    {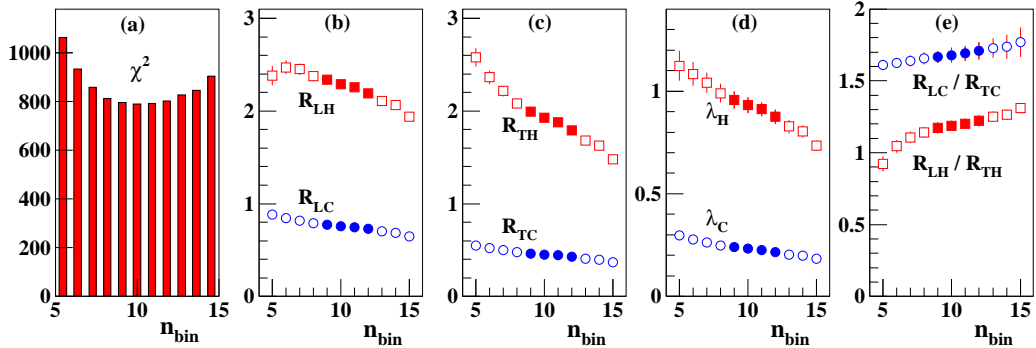}
\end{center}
\caption{Dependence on $q_{\rm cut}$. (a) Combined $\chi^2$ for core
  and halo fits as a function of $q_{\rm cut} = (0.02\, n_{\rm
    bin})\,$GeV.  The four lowest $\chi^2$ values correspond to
  $q_{\rm cut} = 0.18{-}0.24\,$GeV and $\chi^2/\mathrm{NDF} = 1.28$.
  (b)--(e): Dependence of parameter values on $q_{\rm cut}$.  Filled
  points correspond to the four smallest $\chi^2$ in (a).}
\end{figure}
\begin{figure}[htb]
\includegraphics[width=130mm,bb=55 30 530 200,clip=] 
    {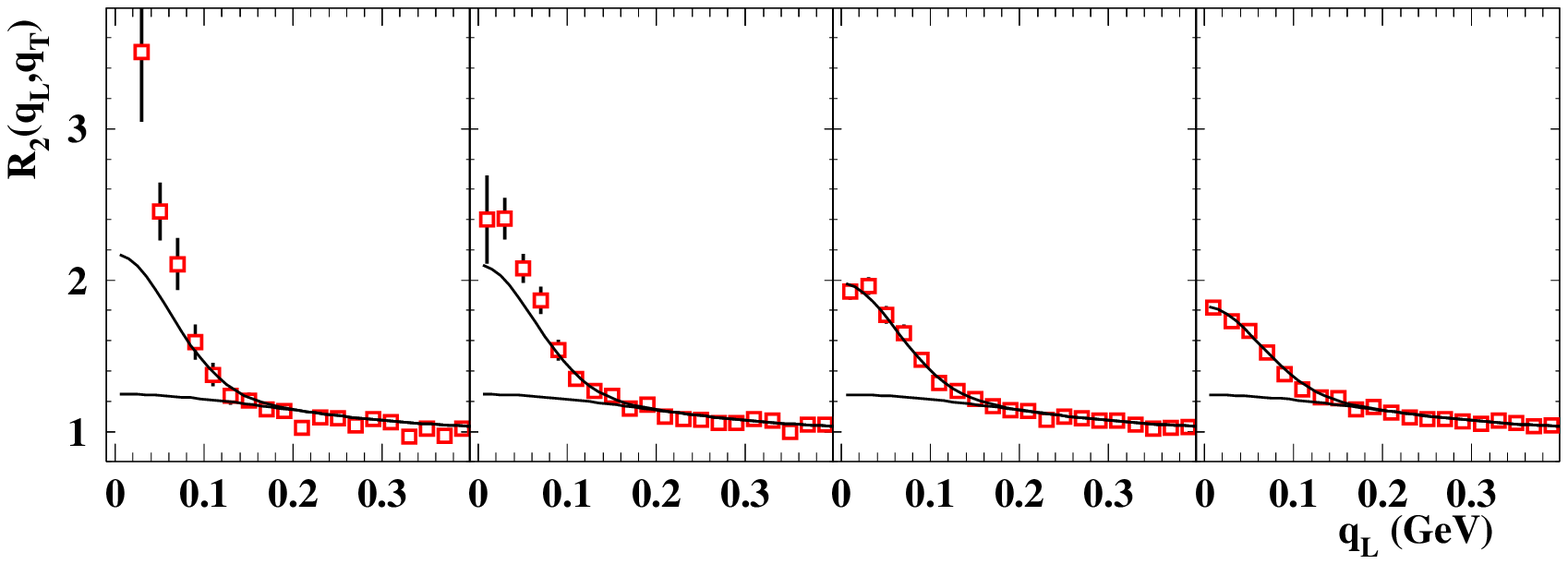}\\[1mm]
\includegraphics[width=130mm,bb=55 30 530 200,clip=] 
    {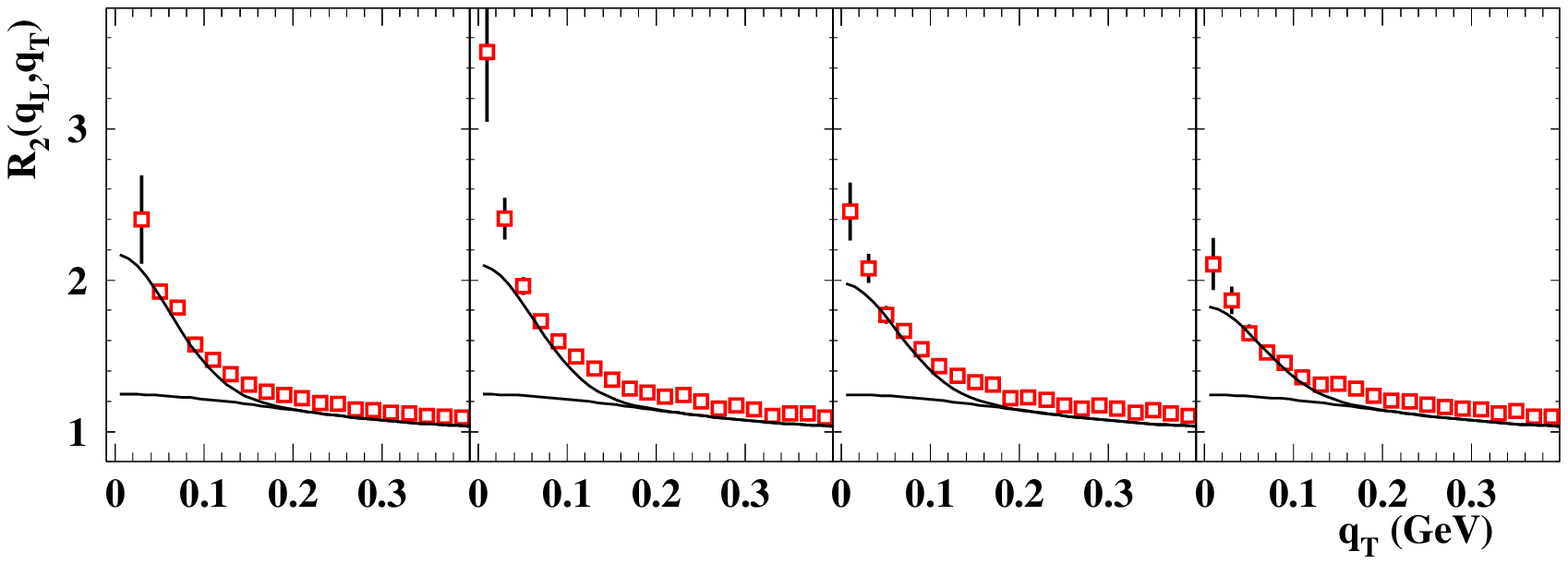}\\[-5mm]
    \caption{$R_2(q_L,q_T)$ and fit to core (lower lines) and combined
      core-halo (upper lines), shown in the upper panels for slices
      with fixed $q_T =$ 0.00--0.02, 0.02--0.04, 0.04--0.06 and
      0.06--0.08 GeV and in the lower panels for corresponding
      fixed-$q_L$ slices.}
      \ \\[-5mm]
\end{figure}
\begin{figure}[htb]
\includegraphics[width=130mm,bb=55 30 530 200,clip=] 
    {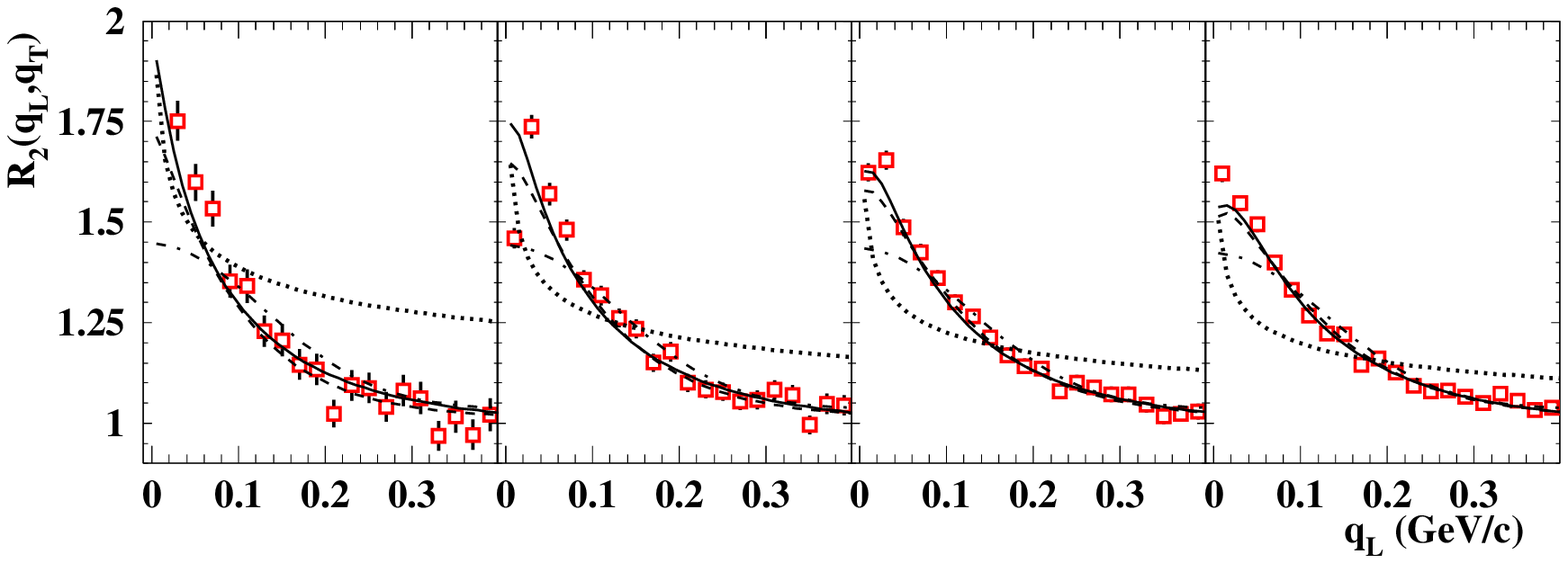}\\[1mm]
\includegraphics[width=130mm,bb=55 30 530 200,clip=] 
    {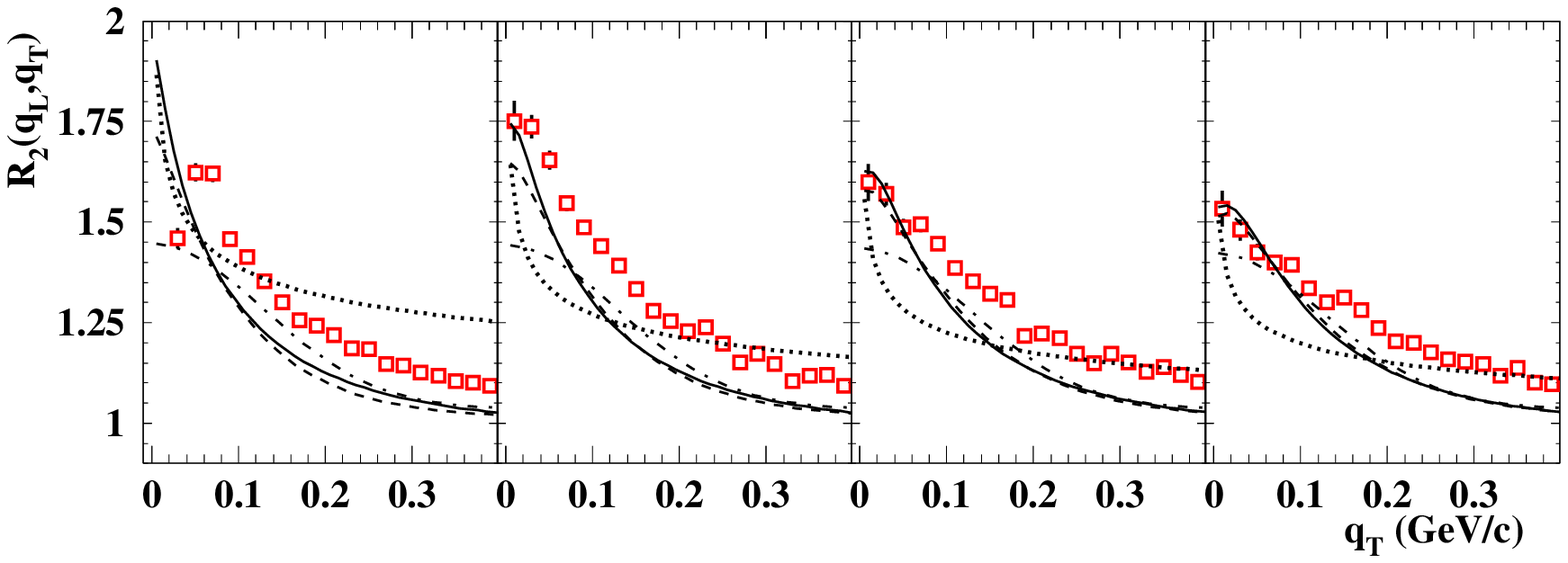}\\[-5mm]
\caption{Same as Fig.~1 but for uncorrected data and including power
  law (dotted line). Note the different vertical scale.}
\end{figure}

\end{document}